\documentclass[9pt,preprint]{aastex}

\begin{document}


\title{Wind-Driven Evolution of White Dwarf Binaries to Type Ia Supernovae}
\author{Iminhaji Ablimit\altaffilmark{1,2},
Xiao-jie Xu\altaffilmark{1,2} and X.-D Li\altaffilmark{1,2}}
\altaffiltext{1}{Department of Astronomy, Nanjing University, Nanjing 210093, China}
\altaffiltext{2}{Key Laboratory of of Modern Astronomy and Astrophysics,
Ministry of Education, Nanjing 210093, China}


\begin{abstract}
In the single degenerate scenario for the progenitors of type Ia
supernovae (SNe Ia), a white dwarf rapidly accretes hydrogen-
or helium-rich material from its
companion star, and appears as a supersoft X-ray source.
This picture has been challenged by the properties of the supersoft
X-ray sources with very low-mass companions and the observations of several
nearby SNe Ia.
It has been pointed out that, the X-ray radiation or the wind from the
accreting white dwarf can excite wind or strip mass from the companion
star, thus significantly influence the mass transfer processes.
In this paper we perform detailed calculations of
the wind-driven evolution of white dwarf binaries. We present the
parameter space for the possible SN Ia progenitors, and for the surviving
companions after the SNe. The results
show that the ex-companion stars of SNe Ia have characteristics
more compatible with the observations, compared to those in the traditional
single degenerate scenario.

\end{abstract}


\section{Introduction}
Type Ia supernovae (SNe Ia) are among the most luminous explosions
in the observed Universe \citep{Fi1997}. The cosmological precision distance
measurements enabled by SNe Ia first revealed the accelerating expansion of
the Universe (\citealp{Riess1998, Per1999}). General consensus
holds that SNe Ia  arise from the thermonuclear explosions of carbon-oxygen white
dwarfs (CO WDs) in close binaries (\citealp{Hoyle1960, Nomoto1982, Iben1984, Nomoto1997}).

The mainstream progenitor models of SNe Ia can be categorized as the single-degenerate
 (SD) and the double-degenerate  (DD) scenarios
 (For reviews, see \citealp{Bra1998, Podsiadlowski2010, Wang2012}).
In the SD scenario, a CO WD accretes hydrogen/helium-rich material from its stellar
companion until its mass approaches the Chandrasekhar mass
$M_{\rm Ch}=1.38\, M_\odot$
 \citep{Welan1973, Nomoto1982, Munari1992, Hachisu1996,Li1997, langer00,hp04,it04,L2009,wlh10}.
The DD scenario involves two WDs in a close binary, and the merger of the two WDs,
which have a combined mass larger than or equal to $M_{\rm Ch}$,
may result in the explosion as a SN Ia
 \citep{Tutukov1981, Iben1984, Webbink1984,Yungelson1994}. In addition,
\citet{ks11} suggest that during the final stages of the common envelope (CE) evolution
a merger of a WD with the core of the asymptotic giant branch (AGB) or the red giant branch
(RGB) star  may trigger a SN Ia, which is termed as the core-degenerate scenario.
Observational and theoretical evidence exists both for and against
each of the progenitor scenarios. At present it is unclear whether one of
these possibilities is exclusively realized in Nature or whether all
contribute to SNe Ia \citep[e.g.][]{howell11}.

In this work we focus on the SD scenario, which is observationally manifested by the
supersoft X-ray sources (SSSs). In these sources a WD is thought
to be accreting rapidly from its binary companion \citep{Kah1997}.
\cite{van1992} explained the observed
characteristics of  SSSs with steady nuclear burning of hydrogen on the WDs.
In this model, mass is transferred from a main-sequence (MS) or (sub)giant donor star
on a thermal or nuclear timescale by Roche lobe overflow (RLOF).
For steady hydrogen burning the accretion rate should be  $\gtrsim 1.0\times{10}^{-7}
\,M_\odot{\rm yr}^{-1}$ for a $1M_\sun$ WD.  Binary evolutionary
calculations show that the donor star is either a MS star more
massive than the WD (with mass $\sim 2-3.5\,M_{\odot}$), or a low-mass
($\sim 1\,M_{\odot}$) (sub)giant star
\citep[e.g.][]{Rappaport1994,Hachisu1996,Li1997,Yungelson1996, langer00,wlh10},
to ensure that the mass transfer rate is high enough for steady burning.

A number of observational results have been used to constrain the SN
Ia progenitor models, including the surviving companion stars of SNe
Ia, the signatures of gas outflows from the SNe Ia progenitor
systems, the early multi-wavelength emission of SNe Ia, the stripped
mass of the companions due to SN explosions, etc \citep[see][and
references therein]{Wang2012}. In particular the delay time
distribution (DTD) of SNe Ia presents a useful tool to probe the
progenitor scenarios. From studies using various SN samples,
environments, and redshift ranges, a best fit power-law form
$t^{-1}$ has been derived, in which a similar DTD with prompt (a few
$10^8$ yr), delayed, and intermediate ( a few Gyr) components was
recovered \citep[see][for a review]{Maoz2012}. This requires that in
the SD scenario the donor mass should be in the range of $\lesssim
1\,M_\odot$ to $\sim 7\,M_\odot$, considerably wider than predicted
by the aforementioned investigations. However, it is well known that
when the donor mass is more massive than $\sim (3-4)\,M_\odot$ the
mass transfer is dynamically unstable and a CE stage must follow.
Moreover, the observational properties of some SSSs (e.g., RX
J0439.8$-$6809, 1E0035.4$-$7230, RX J0537.7$-$7034 and CAL 87) do
not fit in the classical picture of SSSs  as being driven by thermal
timescale mass transfer. For example, the high luminosity ($\sim
10^{37}$ ergs$^{-1}$) and low temperature ($\sim 5\times10^{5}$ K)
measured in 1E0035.4$-$7230 establish that it is a supersoft binary
which has a WD with stable nuclear burning. However, the orbital
period of 4.126 hr \citep{Schmidtke1996} is significantly shorter
than that required for a binary with thermal-timescale mass
transfer, and implies a mass ratio $M_2/M_1\lesssim  0.7$ (where
$M_1$ and $M_2$ are the masses of the WD and the donor respectively)
\citep{van1998}. It is also interesting to note that the prototype
SSS CAL 87 has a WD of mass $1.35\, M_\odot$ \citep{Starr2004} in a
10.6 hr orbit \citep{Naylor1989, Callanan1989}.
The donor masses and orbital periods in these
systems are in contradiction with the standard picture of SSSs.

To resolve the above inconsistencies, a wind-driven evolution of
SSSs was proposed in the
literature. There are two kinds of such models in which the ``wind" has different meaning and
operates in different ways. In relatively low-mass systems, \citet{van1998} and
\citet{King1998} suggest that the companion star can be
irradiated by the soft X-rays from the SSS, exciting strong winds,
which drive RLOF at a high rate and sustain steady hydrogen burning on the accreting WD.
For relatively massive systems, \citet{Ha08a} propose that the optically thick winds from
the accreting WD can collide with the companion and strip off its surface layer.
This mass-stripping attenuates the mass transfer rate, thus preventing the formation
of a CE. \citet{Ha08b} showed that the predicted DTD from $t\sim 0.1-10$ Gyr
is in agreement with the recent measurements.

It is noted that in their studies, \citet{van1998}, \citet{King1998}, and \citet{Ha08a} used
a semi-analytical method or semi-empirical input to investigate the evolution
of the WD binaries. In this paper we present detailed numerical calculations
of the mass transfer processes in the WD binaries including the wind effect.
In section 2 we introduce
the two kinds of wind-driven models. The calculated results are presented
in section 3 and compared with observations in section 4.
Discussion and conclusions are in section 5.


\section{The Wind-driven models}
\label{sec:model}
We calculate the evolution of a binary consisting of a WD and a MS companion star
with an updated version of Eggleton's stellar evolution
code \citep{Eggleton1971, Eggleton1973}. We take into account the self-excited
winds from the donor due to the irradiation from the WD \citep{King1998} (termed as case 1),
the mass-stripping from the donor due to the WD's wind \citep{Ha08a}
(termed as case 2), and the response
of the donor due to the related mass loss processes. Beside angular momentum
losses due to the winds or the stripped material, angular momentum loss
caused by gravitational wave radiation \citep{ll75} and magnetic braking
\citep{vz81,r83} is also considered.
Below we introduce the self-excited wind model and the mass-stripping model.

\subsection{The self-excited wind model}

\citet{van1998} argue that irradiation in perhaps all SSSs may lead to a strong
stellar wind from the heated side of the donor star. If
escaping the binary with the specific angular momentum
of the donor, the wind will drive mass transfer with a rate
which is of the same order as the wind loss rate.
The relation between the  mass transfer rate $\dot{M}_{\rm tr}$
and the wind loss rate $\dot{M}_{\rm w}$ obeys
\begin{equation}
\dot{M}_{\rm w}\simeq 3.5\times 10^{-7}\,M_\sun\,{\rm yr}^{-1}
(\frac{M_2}{M_\sun})^{5/6}(\frac{M}{M_\sun})^{-1/3}
(\eta_{\rm s}\eta_{\rm a})^{1/2}\phi
(\frac{\dot{M}_{\rm tr}}{10^{-7}\,M_\sun\,{\rm yr}^{-1}})^{1/2},
\end{equation}
for $M_2\lesssim M_1$; and
\begin{eqnarray}
\dot{M}_{\rm w}\simeq 3.5\times 10^{-7}\,M_\sun\,{\rm yr}^{-1}
(\frac{M_2}{M_\sun})^{0.95}(\frac{M}{M_\sun})^{-1/3}
(\frac{M_1}{M_\sun})^{-0.12}(\eta_{\rm s}\eta_{\rm a})^{1/2}\phi
(\frac{\dot{M}_{\rm tr}}{10^{-7}\,M_\sun\,{\rm yr}^{-1}})^{1/2},
\end{eqnarray}
for $M_2\gtrsim M_1$. Here $M=M_1+M_2$, $\eta_{\rm s}$ measures the efficiency of
the WD's spectrum in producing ionizing photons normalized to the case of supersoft
X-ray temperatures of a few times $10^5$ K, $\eta_{\rm a}$ measures
the luminosity per gram of matter accreted relative to the value for hydrogen
shell burning, and $\phi$ is an efficiency factor
parameterizing the fraction of the companion's irradiated face, and the fraction
of the wind mass escaping the system \citep[for the details, see][]{King1998}.
Stability analysis
shows that, for SSSs with a low mass ratio ($\lesssim 1$),
the resulting mass transfer is stable and with a rate sufficient
to keep the binary in the stable (steady or recurrent) hydrogen
burning regime; for WD binaries with large
mass ratios ($\gtrsim 1$), the self-excited winds can stabilize mass
transfer at a threshold value ($\sim 10^{-8}\,M_\sun$yr$^{-1}$)
for nonexplosive nuclear burning of the accreted matter \citep{King1998}.

\subsection{ The mass-stripping model}
\citet{Ha08a} introduce the mass-stripping effect on a MS
or slightly evolved companion star by the winds from an accreting WD.
Since the companion star is initially more massive than the WD, mass transfer
proceeds on a thermal timescale. If the mass transfer rate exceeds the critical
rate  \citep[e.g.,][]{Nomoto1982},
\begin{equation}
\dot{M}_{\rm cr}=7.2\times10^{-6}({M_{\rm WD}/M_\odot}-0.6)\,  M_\odot\rm yr^{-1},
\end{equation}
there is optically thick wind to blow from the
WD \citep{Hachisu1996}. The fast-moving wind collides with the companion star
and strips off its surface layer by the shock dissipated into it \citep{Ha03}.
The stripping rate is estimated to be
\begin{equation}
\dot{M}_{\rm strip}=C_1\dot{M}_{\rm wind},
\end{equation}
where $C_1$ is a numerical factor determined by the binary separation
and the masses of both components with its value ranging $\sim 1-10$,
and $\dot{M}_{\rm wind}$ is the wind loss rate from the WD.
The stripped mass is assumed to leave the binary from the $L_3$ point,
carrying away the corresponding specific angular momentum.  Since this mass-stripping
quickly decreases the mass of the donor star, it can
attenuate the mass transfer rate, thus preventing the formation of a CE.
In this way, young and massive companion stars can drive the WDs to the SN Ia explosions.

\subsection{Growth of of the WD mass}

The growth rate of the WD in mass as a result of the accretion
is determined by the accumulation efficiencies of hydrogen and helium burning,
\begin{equation}
\dot{M}_{\rm WD}=\eta_{\rm H}\eta_{\rm He}\dot{M}_{\rm tr},
\end{equation}
where $\eta_{\rm H}$ and $\eta_{\rm He}$ represent
the fraction of transferred hydrogen and helium-rich matter from the companion
that eventually burns into helium and carbon-rich matter and stays on
the WD, respectively. Here we fit the numerical results
of \citet{Pralnik1995} and \citet{Yaron2005} for the hydrogen mass accumulation efficiency
$\eta_{\rm H}$, and adopt the prescriptions in \citet{Ka2004} for the helium mass accumulation
efficiency $\eta_{\rm He}$.

\section{Results}
We have calculated the evolutions of a grid of binaries with WDs of
initial mass $M_{\rm WD,i}=1.0\,M_\odot-1.2\, M_\odot$ and  MS companion stars
of initial mass
$M_{\rm 2,i}=0.6\, M_\odot-7\, M_\odot$ with Solar abundance in cases 1 and 2, respectively.

Figure 1 shows the distribution of the initial companion's mass $M_{\rm 2,i}$ versus
the initial orbital period $P_{\rm orb,i}$ in which the binaries can evolve to  SNe Ia in case 1.
The black, red, and blue solid lines correspond to the initial WD mass of $1\,M_\sun$,
$1.1\,M_\sun$, and $1.2\,M_\sun$, respectively. The dashed lines
show the results with the same WD mass but without self-excited winds considered,
as calculated by \citet{Li1997}. In the latter case the mass transfer is
driven only by the nuclear evolution of the companion star. To acquire a sufficiently high mass
transfer rate the companion star  should be more massive than $\sim 2\,M_\odot$.
When the self-excited wind from the companion star is included, the companion
star's mass can extend down to $\lesssim 1\,M_\odot$.
Since the wind can cause
the orbit to shrink and accelerate the mass transfer,
the mass transfer can maintain a rate high enough
for stable burning, even if $M_2$ is not
larger than $M_1$. However, this effect also lowers the upper limit of the
companion's mass and narrows the initial orbital period to some extent, because the mass transfer
becomes dynamically unstable if $M_{\rm 2,i}$ or $P_{\rm orb,i}$ is larger.

The initial $M_{\rm 2,i}$ vs. $P_{\rm orb,i}$ distributions in case 2 (with $C_1=3$)
are shown in Figure 2a.
Figure 2b shows the same distribution but with fixed $M_{\rm WD,i}=1.2\,M_\odot$
and different values of $C_1$ (the ``no-wind" result is also presented).
Compared with \citet{Ha08a}, there are similar features such as,
the allowed range of the companion star's mass
is larger for bigger $C_1$. However, the parameter spaces in Figure 2 is
somewhat smaller than in \citet{Ha08a}, because in \citet{Ha08a} the boundaries of
the SN Ia region in the $M_{\rm 2,i} - P_{\rm orb,i}$
plane are set by the following simplified conditions:
(1) The left boundary is given by the mass-radius relation
for the zero-age MS stars; (2) the lower boundary is
set by the occurrence of strong nova explosions, when
$\dot{M}_{\rm tr}\lesssim 10^{-7}\,M_\sun$ yr$^{-1}$;
 (3) the upper boundary is set by the formation of a CE
with the assumption of  $\dot{M}_{\rm tr}\gtrsim 10^{-4}\,M_\sun$ yr$^{-1}$; and
 (4) the right boundary corresponds to the end
of central hydrogen burning of the MS companion.
Additionally, \citet{Ha08a} use a semi-analytical approximation to estimate
the thermal-timescale mass transfer rate.
In our calculations, although the mass-stripping can result in a lower mass transfer
rate (compared with the no-stripping case), the system still has a too high
mass transfer rate and evolves into the CE phase if the companion star's
mass $M_{\rm 2,i}\gtrsim 4-6\,M_\odot$.
For lower-mass companion stars ($M_{\rm 2,i}\lesssim 2-3\,M_\odot$) the mass transfer
is not fast enough to drive the WD to the Chandrasekhar mass.

Comparison of Figures 1 and 2 shows that the SN Ia
regions in case 2
are larger than in case 1, but the lower limits of $M_{\rm 2,i}$ are not as low as
in case 1. Although both the irradiation and the wind from the WD can trigger
mass loss from the companion star, their influence on the orbital evolution
and hence the mass transfer is different. We find that angular momentum loss
and mass loss dominate the evolution in cases 1 and 2 respectively, so that
the mass transfer rates for low/high-mass companion stars are higher/lower
in case 1 than in case 2.
Therefore, in general, the mass-stripping effect favors more massive
companion stars to drive the WDs to SN Ia explosions,
while the self-excited wind is preferred for lower-mass companion stars.

Figure 3 compares the distributions of the companion mass $M_{\rm 2,f}$ and
the orbital period $P_{\rm orb,f}$ when the WD reaches $1.38\,M_\odot$ in cases 1
(left) and  2 (right). While the orbital periods are distributed
similarly, e.g., between $\sim 1$ day and $\sim 12-14$ days
for $M_{\rm WD,i}=1.2M_\sun$, the masses of the companion stars in case 1 range from
$\sim 0.3M_\sun$ to $\sim 1.8M_\sun$, significantly smaller than in case 2
($0.3M_\sun\lesssim M_{\rm WD,f}\lesssim 4.5 M_\sun$). In both cases
lower-mass companion stars are more likely to be in wider orbits
because of longer mass transfer time.

In Figures 4-6 we show the distributions of the companion's rotating
velocity
vs. the stellar radius, the surface gravity vs. the orbital
velocity, and the effective temperature vs. the absolute
magnitude when the WD mass reaches $1.38 M_\sun$ in
cases 1 (left) and 2 (right), respectively. Together with Figure 3, they present
useful probes to test the proposed SD scenarios by comparison with
observational constraints \citep[cf.][]{han08}\footnote{Note that
Figures 3-6 are for the moment of the SN explosion;
the realistic distributions could be changed due to the explosion.
Furthermore, the donor star might evolve after the mass transfer but before the SN
explosion, possibly due to the spin-down of the WD to reach the critical
density in its core \citep{Justham2011,Stefano2011}.}.

\section{Comparison with observations}
\subsection{The surviving companions after the SNe Ia}
The Tycho's supernova (SN 1572) is close enough to
search for the widowed companion to the exploded WD, should it exist.
 \citet{Ruiz2004} surveyed the central region of SN 1572's remnant,
and proposed that the star G appears to be the surviving companion
of the SN. This star is similar to our Sun in surface temperature
($5750\pm 250$ K) but with higher space velocity of $\sim 136$ kms$^{-1}$
and lower surface gravity ($\sim 0.1-1.0$ km\,s$^{-2}$). These values
are within the allowed regions presented in Figures 5 and 6.
However, more recent observations  indicate that this identification
remains controversial \citep{Fuhrmann2005,Ihara2007,Gonz2009,Ker2013b}.

\citet{Li2011} used the extensive historical imaging obtained at the location of SN
2011fe/PTF11kly, the closest SN Ia discovered in the digital imaging era, to constrain
the visible-light luminosity of the progenitor to be $10-100$ times fainter than previous
limits on other SNe Ia progenitors. They proposed that the exploding WD
accreted matter either from another WD, or by RLOF from a subgiant or
MS companion star.
\citet{Edwards2012} searched for the ex-companion star in SNR 0519$-$69.0,
located in the Large Magellanic Cloud, based on the Hubble Space Telescope (HST) images
with a limiting magnitude of $V = 26.05$.  They found that one of the 27
MS stars brighter than $V = 22.7$  (corresponds to $M_V = 4.2$)
within $4\farcs7$ of the position
could be the ex-companion star left over from a SSS progenitor, ruling out SD models
with post-MS stars.
\citet{Ker2013a}
spectroscopically scrutinized 24 of the brightest stars residing in the central
$38\arcsec \times 38\arcsec$ of the SN 1604 (Kepler) SN remnant to search for a
possible surviving companion star, and ruled out donor star down to $10L_\sun$.
\citet{Gon2012}  searched for the surviving companions of
the progenitor of
SN1006, and found that none of the stars within $4\arcmin$
of the apparent site of the explosion is associated with the SN remnant
down to the limit $M_V\simeq 4.9$, thus exclude all giant and subgiant companions
to the progenitor.
So SN 1006 should have been produced either by mass accretion from an unevolved star,
similar to, or less massive than the Sun, or by merging with another WD.
\citet{Ker2012} also scrutinized the central stars
(79 in total) of the SN 1006 remnant to search for the surviving donor star.
The rotational velocities of these stars are estimated to be around 10 kms$^{-1}$.
The aforementioned observational constraints on the possible surviving companions
deviate significantly from expected in the traditional SD models,
but are (in some cases barely) consistent with the results in the self-excited
wind model (case 1) shown in Figures 4-6.

From the HST deep images of SNR 0509$-$67.5 in
the Large Magellanic Cloud,
\citet{Schaefer2012} reported that the maximal central error circle is empty of point
sources to the limit of $V = 26.9$, which corresponds to $M_V = 8.4$.
This is outside our predicted region even in the most optimistic
situation. However, as discussed below,  lower-mass
companion stars are expected,
if other effects such as the WD's magnetic field are taken into
account.

\subsection{The delay time distribution}
Another useful feature of the SN Ia progenitors is the DTD.
The DTD is the hypothetical SN rate versus time that would follow a
brief burst of star formation, and
the delay times are defined as the time interval
between the star formation and the SN explosion. The SN
Ia DTDs are directly linked to the lifetimes
of the WD's progenitor and the companion star, and the binary
separation. Various kinds of the
progenitor models of SNe Ia can be examined by comparing
their DTDs with observations.
Similar as in \citet{Ha08b}, the DTD of SNe Ia from the
wind-driven systems is estimated by integrating the initial sets of $(M_1, q, a)$
(where $a$ is the binary separation) having the delay time between
$t-\Delta t$ and $t+\Delta t$,
\begin{equation}
{\rm DTD}(t)\propto\frac{1}{2\Delta t}\int\int\int_{t-\Delta t}^{t+\Delta t}
\frac{dM_{1,i}}{M_{1,i}^{2.5}}f(q)dqd\log a.
\end{equation}
Assuming $f(q)=1$ we show the calculated DTDs of the two wind-driven models in Figure 7
(the SN rate has been normalized to be $10^{-3}$ yr$^{-1}$).
Also plotted in the solid line is the fitted form for the DTDs
from current observations \citep{Maoz2012},
\begin{equation}
\Phi(t)=4\times10^{-13} {\rm SN\, {yr}^{-1}} M_\odot^{-1}{(\frac{t}{1\, \rm Gyr})^{-1}},
\end{equation}
where $t$ is the delay time. Figure \ref{fig:hDTD} shows that the
calculated results in both cases are in agreement of Eq.~(7), and those
in case 1 are generally longer than case 2, because of the lower donor mass.
If both models work, the DTDs can extend from a few $10^8$ yr
to $>12$ Gyr.

\section{Discussion and conclusions}

In the standard picture of the SD scenario for SNe Ia, a WD accretes at a rate
$\gtrsim 10^{-7}\,M_\sun$yr$^{-1}$ from its companion star
which is required to power steady hydrogen burning.
The fast accretion onto the WD is driven by the Roche lobe shrinking
(or increase) faster (or more slowly) than
the companion star, and requires a mass ratio of $>5/6$,
that is, the companion stars to $1.38 M_\sun$ WDs
must be either MS stars or (sub)giants  with mass
$>1.16 M_\sun$ \citep{Schaefer2012}.
On the hand, the companion's mass
must be $\lesssim 3-4M_\sun$ to guarantee that the mass transfer
is dynamically stable \citep{Li1997}. This value is obtained
under the assumption
that there are strong winds from the WD if the accretion rate is
higher than the upper limit of the accretion rate for stable hydrogen
burning \citep{Hachisu1996}. The properties of the surviving stars
after the SN Ia explosions can thus been investigated \citep[e.g.][]{han08}.

However, observations of SNe Ia and SSSs, especially the derived DTDs
of SNe Ia, the existence of  1E 0035.4$-$7230-like sources,
and the constraints on the surviving companion stars for several nearby SNe Ia,
strongly suggest that
the initial masses of the companion stars should occupy a wider range
than thought before.
Further revisions on the SD scenarios consider the effects of irradiation
the soft X-rays from the WD on the companion star \citep{van1998,King1998},
a possible
circumbinary disk around the WD binary \citep{Chen2007},
the mass-stripping of the companion star by the WD's wind \citep{Ha08a},
and the thermal instability in the accretion disk \citep{Xu2009,wlh10}.

In this paper, we calculate the evolution of WD/MS star binaries
based on the self-excited wind model \citep{van1998,King1998}
and the mass-stripping model \citep{Ha08a} to produce SNe Ia.
In these models, X-ray radiation or the wind from the accreting
WD strongly affects the structure of the companion's envelope,
causing winds or mass loss from the donor star.
This wind-driven evolution can stabilize the mass transfer when the companion
is considerably more massive than the WD, thus avoiding the
CE evolution, or enhance the mass transfer rate when the companion
is less massive than the WD.
It is shown that the predicted properties of the companion
stars are more compatible with recent observational constraints
for several SNe Ia which have been investigated in detail
than in the traditional SD scenario.

We caution that several issues in the wind-driven evolution models
need to be addressed.
It is unclear whether the self-excited wind and the mass-stripping
work jointly or independently.
The stability and efficiency of
the mass transfer under these circumstances depend on the parameters
$\eta$, $\phi$ and
$C_1$, which are related to the efficiency of X-ray irradiation,
the WD's wind velocity, and the structure of the companion
star, all poorly known.
Furthermore, the conditions for the wind-driven evolution are
not well understood, although they are theoretically permitted.
For example, most of the known WD binaries with low-mass companions are ordinary
cataclysmic variables (CVs) rather SSSs, suggesting that the wind-driven
case may not be popular and some kind of accident is required to trigger the high
luminosities\footnote{One possible way is a long phase of residual hydrogen
burning after a mild shell flash, or a late helium shell flash of
the cooling WD,
after the system has already come into contact as a CV \citep{King1998}.}.
However, there are observational hints that rapid mass transfer
might have occurred in CVs. For example,
\citet{Zorotovic2011} show that the mean WD mass among CVs
($M_{\rm WD} = 0.83 \pm 0.23 M_\sun$) is
significantly larger than that found for pre-CVs ($M_{\rm WD} = 0.67 \pm 0.21 M_\sun$)
and single WDs
\citep[see also][]{Warner1973,Ritter1985,Knigge2006,Yuasa2010, Savoury2011}.
One plausible explanation is that the WD mass
has grown in CVs through accretion during episodes
of stable hydrogen and helium burning.

A potentially important factor we have neglected is the magnetic
field of the WDs. It is well known that a considerable fraction
($\sim 25\%$) of CVs possess magnetic
WDs named as polars and intermediate polars. Recent observations show
that  WDs with magnetic fields $B\lesssim 3$ MG have mean mass
$M_{\rm WD} = 0.68 \pm 0.04M_\sun$; for WDs with $B > 3$ MG,
$M_{\rm WD} = 0.83 \pm 0.04M_\sun$ \citep{Kepler2013}, suggesting
that mass growth may be related to the magnetic field strength.

Hydrogen burning on magnetic and non-magnetic WDs could be in
different ways, even at the same accretion rate.
For accretion onto the entire surface of a WD, the accretion rate
required to sustain stable hydrogen burning is
$\sim 10^{-7}\,M_\sun$yr$^{-1}$. No
CVs can reach such high rates
driven by angular momentum loss due to
magnetic braking and gravitational radiation.
However, if the accreting material is funneled onto a small
portion of the WD surface by the magnetic lines, then the local accretion rate will
be large enough to sustain local stable hydrogen burning
\citep{Schaefer2010}. This enhances
the effective rate of accretion compared to spherical accretion.
The fraction of the WD surface covered by
the accretion spot measured in magnetic CV systems is typically
$\sim 10^{-3}-10^{-2}$ \citep[][and references therein]{Schaefer2010}, thus
steady hydrogen burning requires that
the accretion rate be $\gtrsim 10^{-10}-10^{-9}\,M_\sun$yr$^{-1}$.
This means that magnetic WDs may achieve
a long-lasting SSS phase with a greatly lower accretion rate than
non-magnetic WDs. X-ray irradiation from the
SSS will be able to heat the surface of
the companion star, excite winds and  sustain a relatively high
accretion rate for a long time.
This configuration was recently discussed by \citet{Wheeler2012} in the
context of WD/M dwarf binaries. The author argues that even modest
magnetic fields on the WD and M dwarf will be able to
lock the two stars together, resulting in a slowly rotating WD.
The mass loss will be channeled by a magnetic bottle connecting
the two stars, landing on a concentrated polar area on the WD.
Luminosity from accretion and hydrogen burning
on the surface of the WD may induce self-excited mass transfer.
Obviously the wind-driven evolution for magnetic WDs will be of great
interest as a possible pathway to SNe Ia and be investigated
in the future work.

\acknowledgments
This work was supported by the Natural Science Foundation of China under grant number
11133001 and 11333004, the National Basic Research Program of China (973 Program 2009CB824800),
and the Qinglan project of Jiangsu Province.


\begin{figure}
\includegraphics[totalheight=4.3in,width=5in]{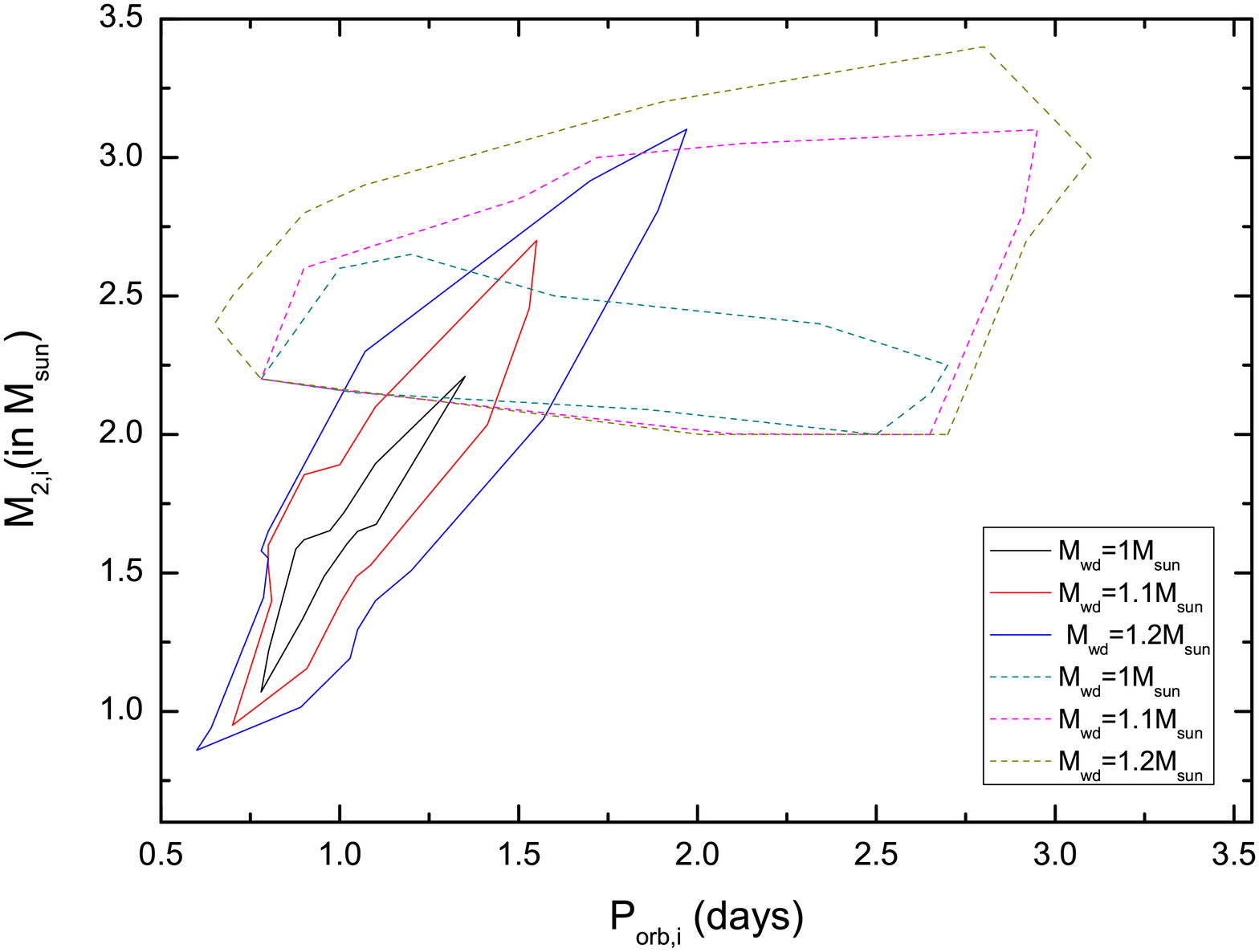}
\caption{The solid lines outline the distributions of the initial orbital periods
and companion masses
of the progenitors of SNe Ia in Case 1. The dashed lines are for no-wind evolution. }\label{fig:1}
\end{figure}

\begin{figure}
\includegraphics[totalheight=3in,width=3in]{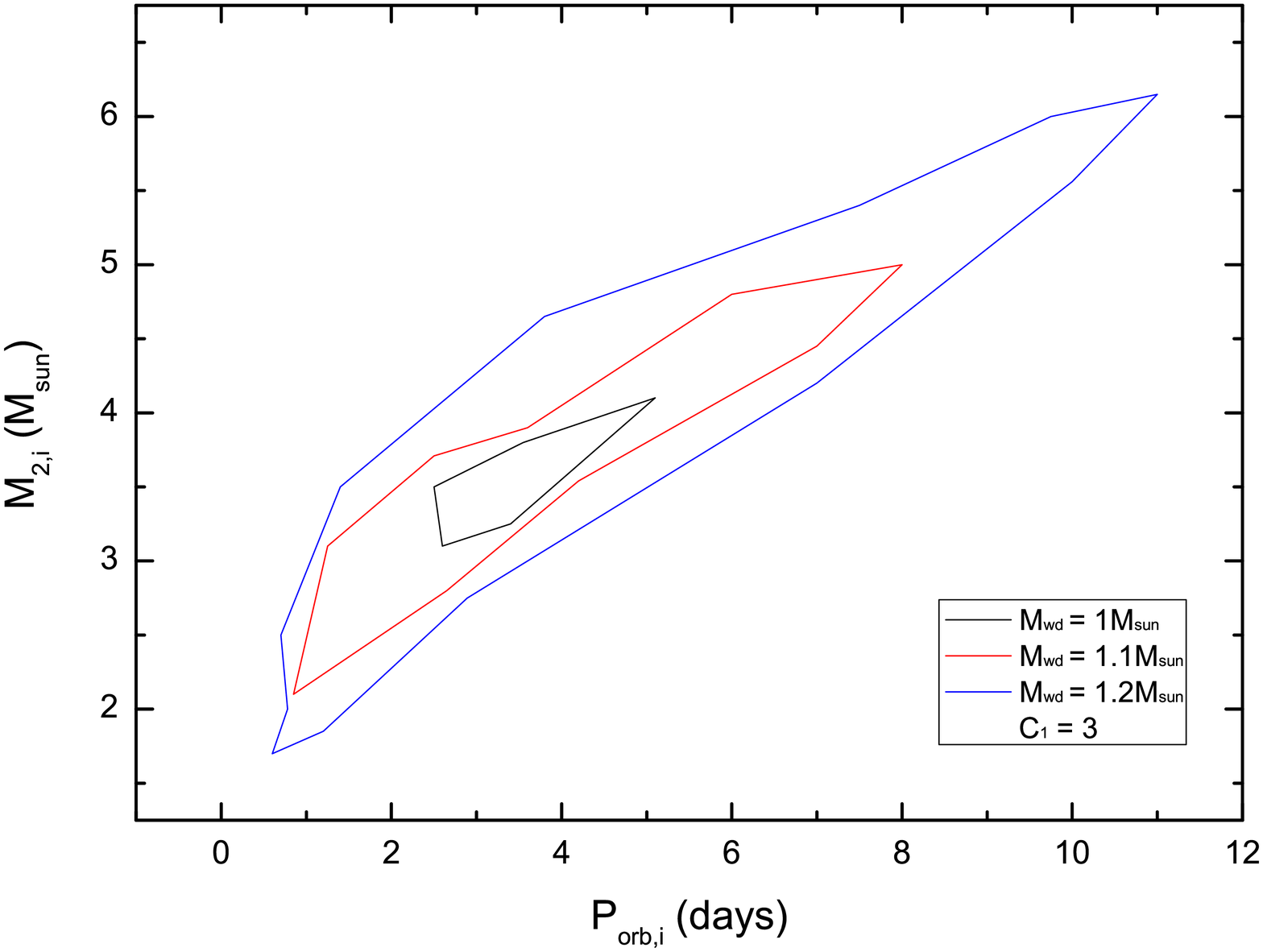}
\includegraphics[totalheight=3in,width=3in]{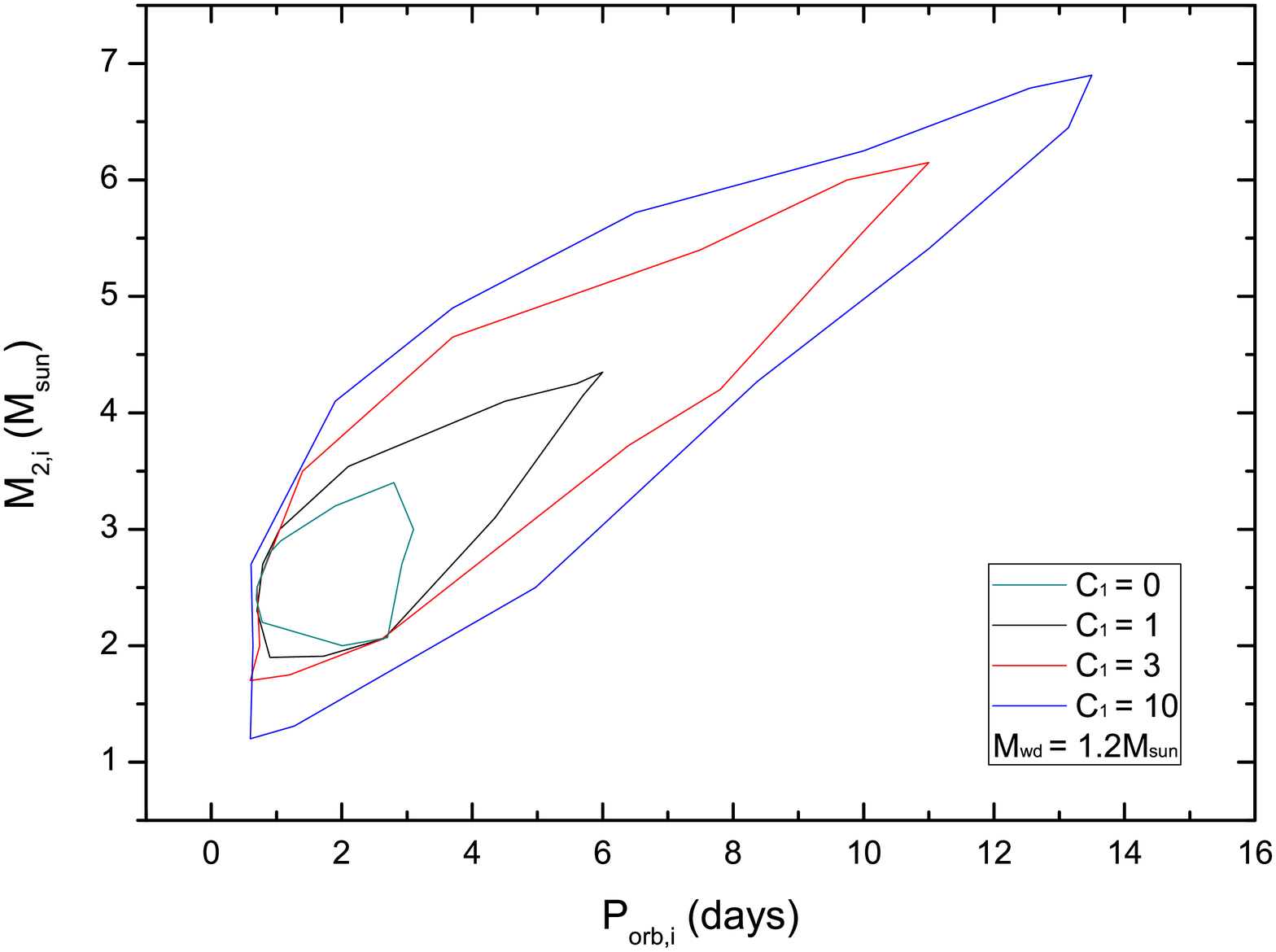}
\caption{Distributions of the initial orbital periods
and companion masses
of the progenitors of SNe Ia in Case 2.}\label{fig:h1}

\end{figure}

\clearpage

\begin{figure}
\includegraphics[totalheight=3.0in,width=3.0in]{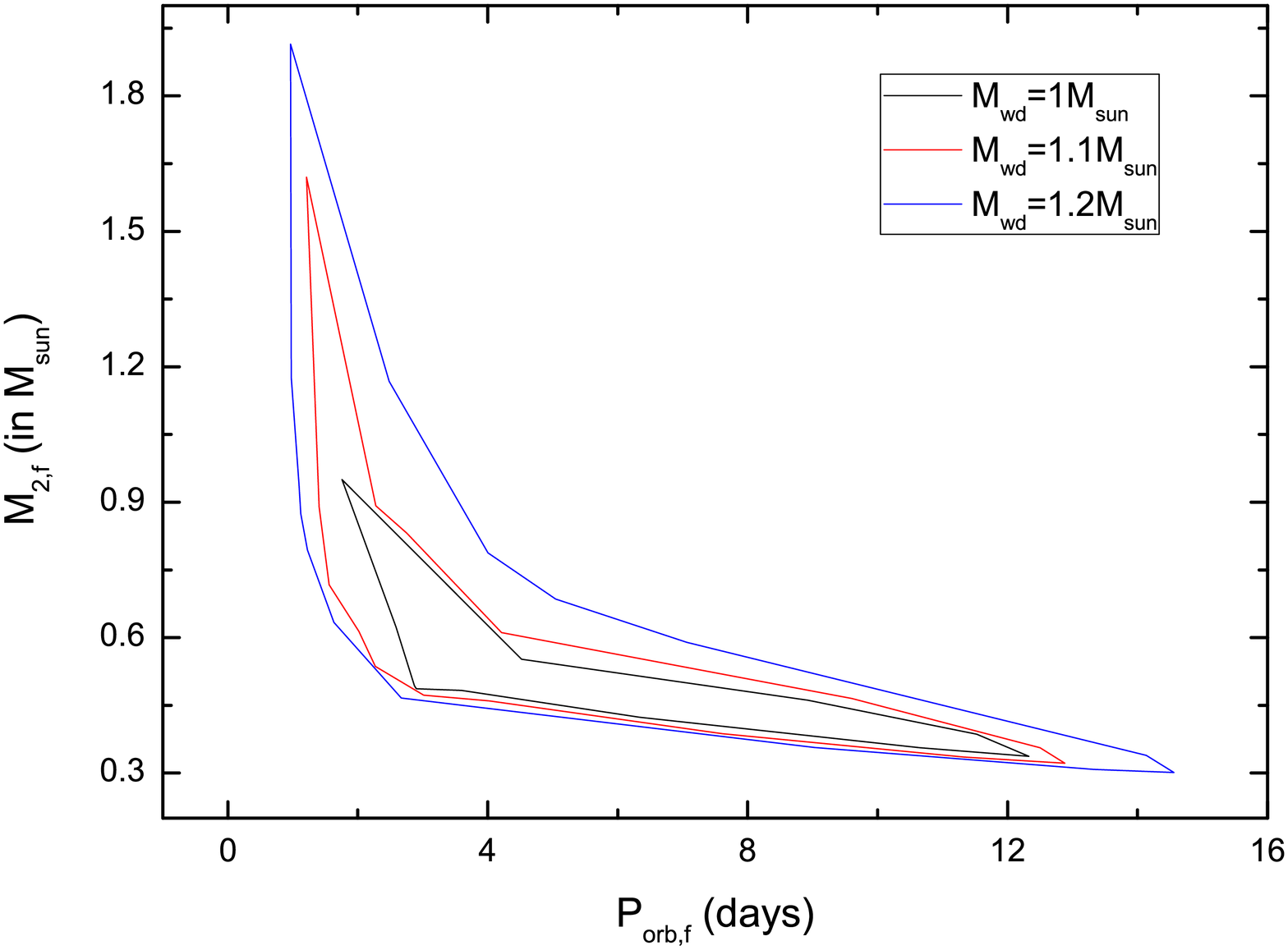}
\includegraphics[totalheight=3.0in,width=3.0in]{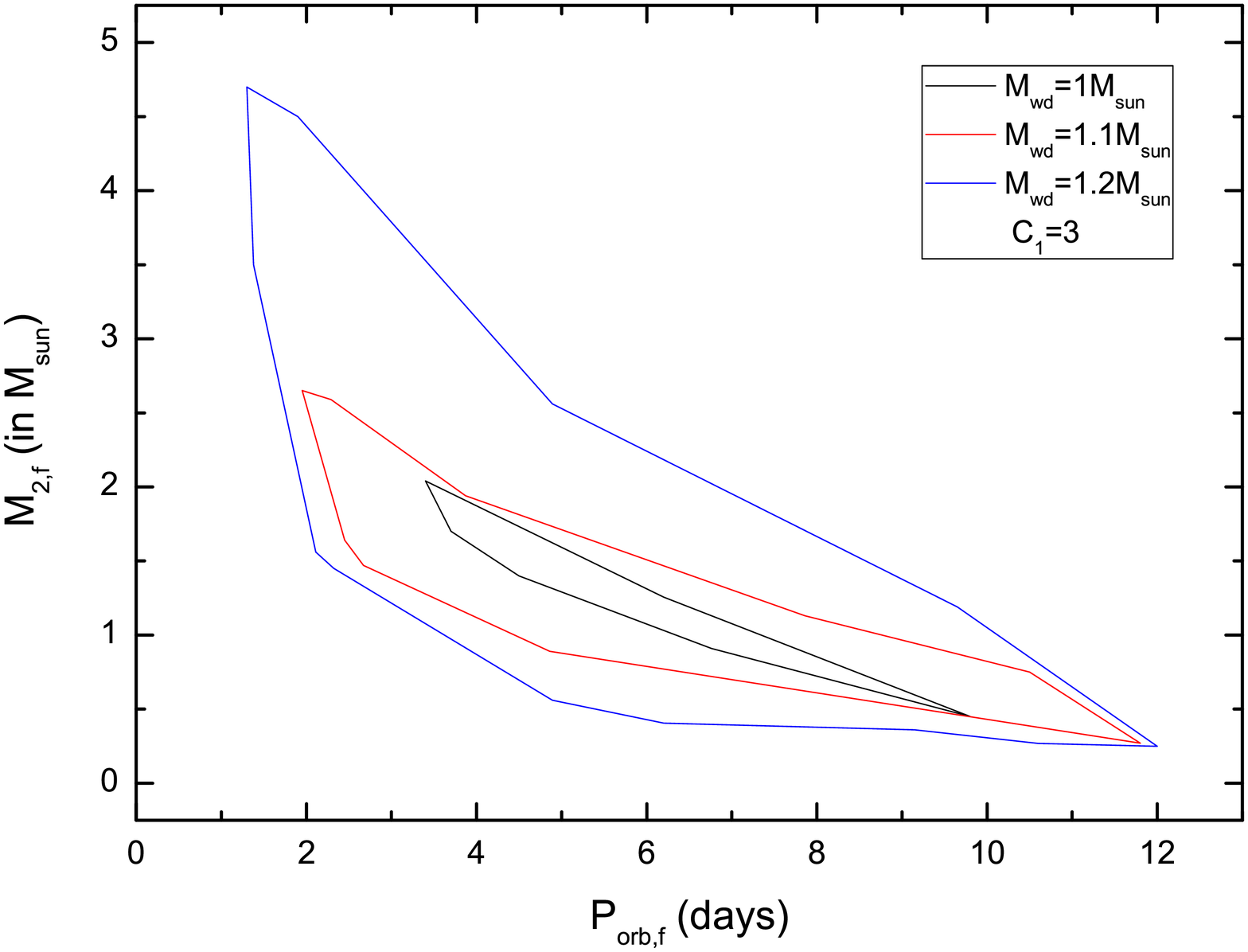}
\caption{ Distributions of the companion masses and orbital periods when the WD grows to
$1.38 M_\odot$ in Cases 1 (left) and 2 (right)}\label{fig:h3}
\end{figure}

\clearpage

\begin{figure}
\includegraphics[totalheight=3in,width=3in]{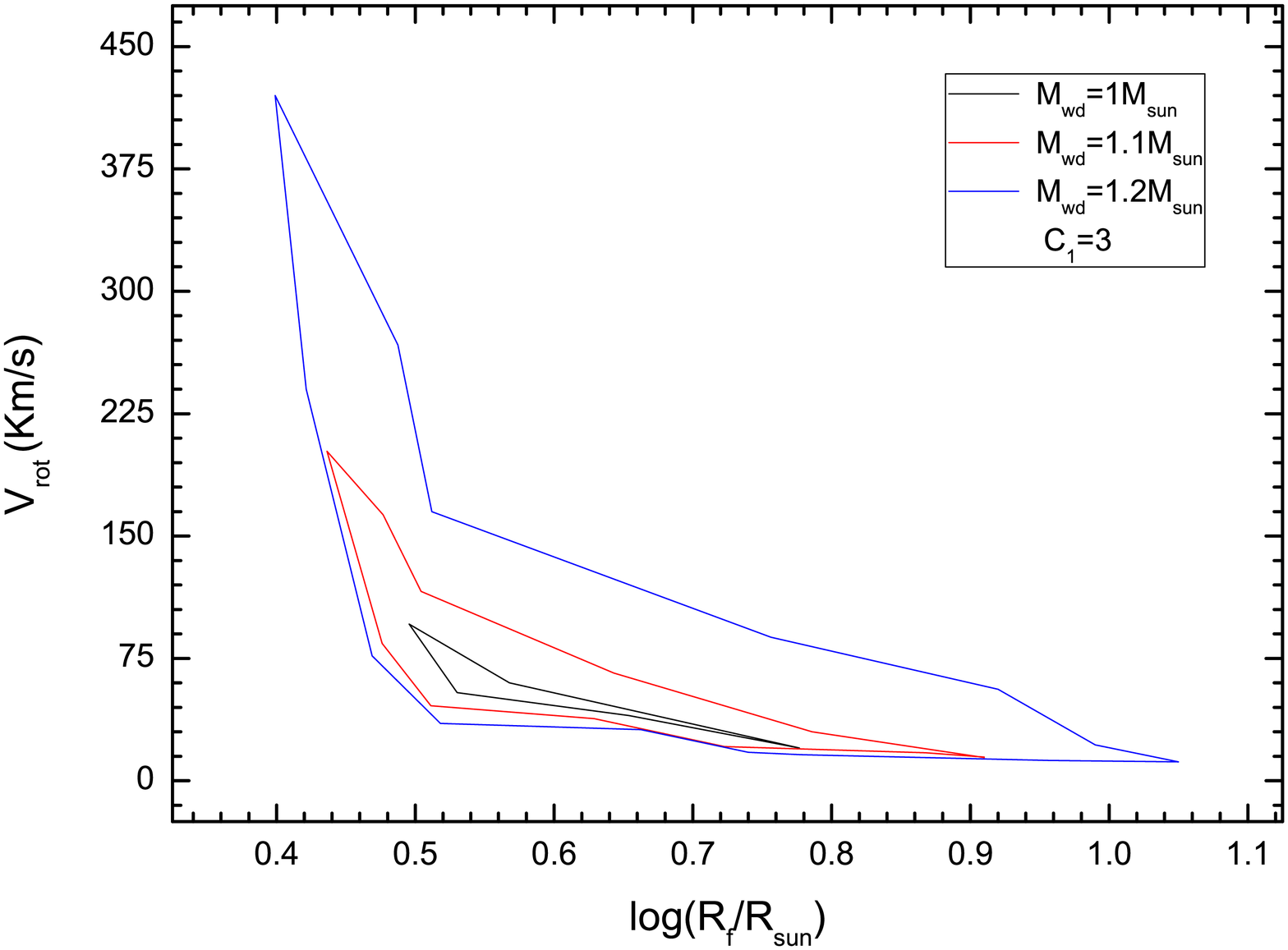}
\includegraphics[totalheight=3in,width=3in]{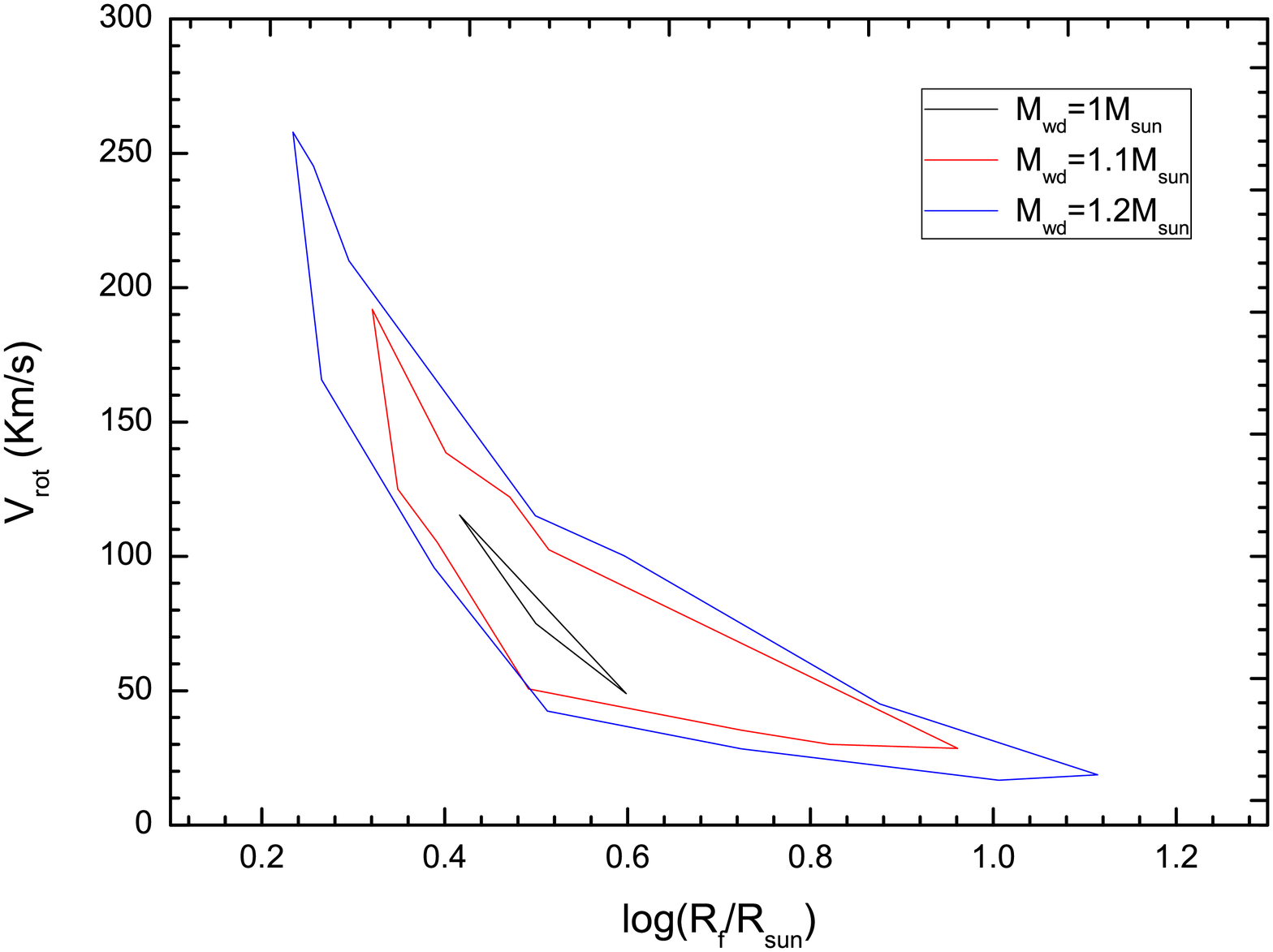}
\caption{Distributions of the rotating velocities and radii of the companion when the WD grows to
$1.38 M_\odot$ in Cases 1 (left) and 2 (right).}\label{fig:h4}
\end{figure}

\clearpage

\begin{figure}
\includegraphics[totalheight=3in,width=3in]{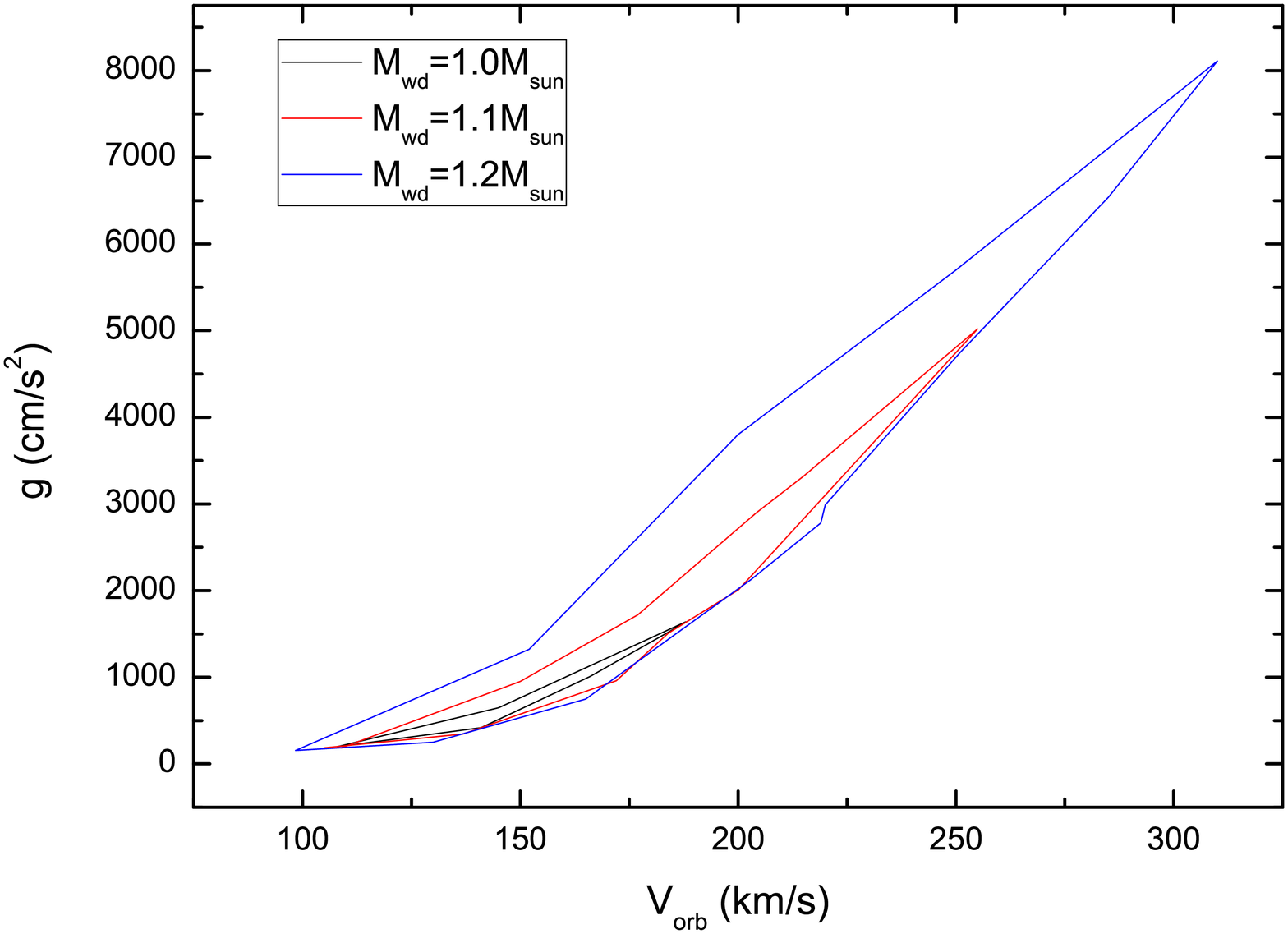}
\includegraphics[totalheight=3in,width=3in]{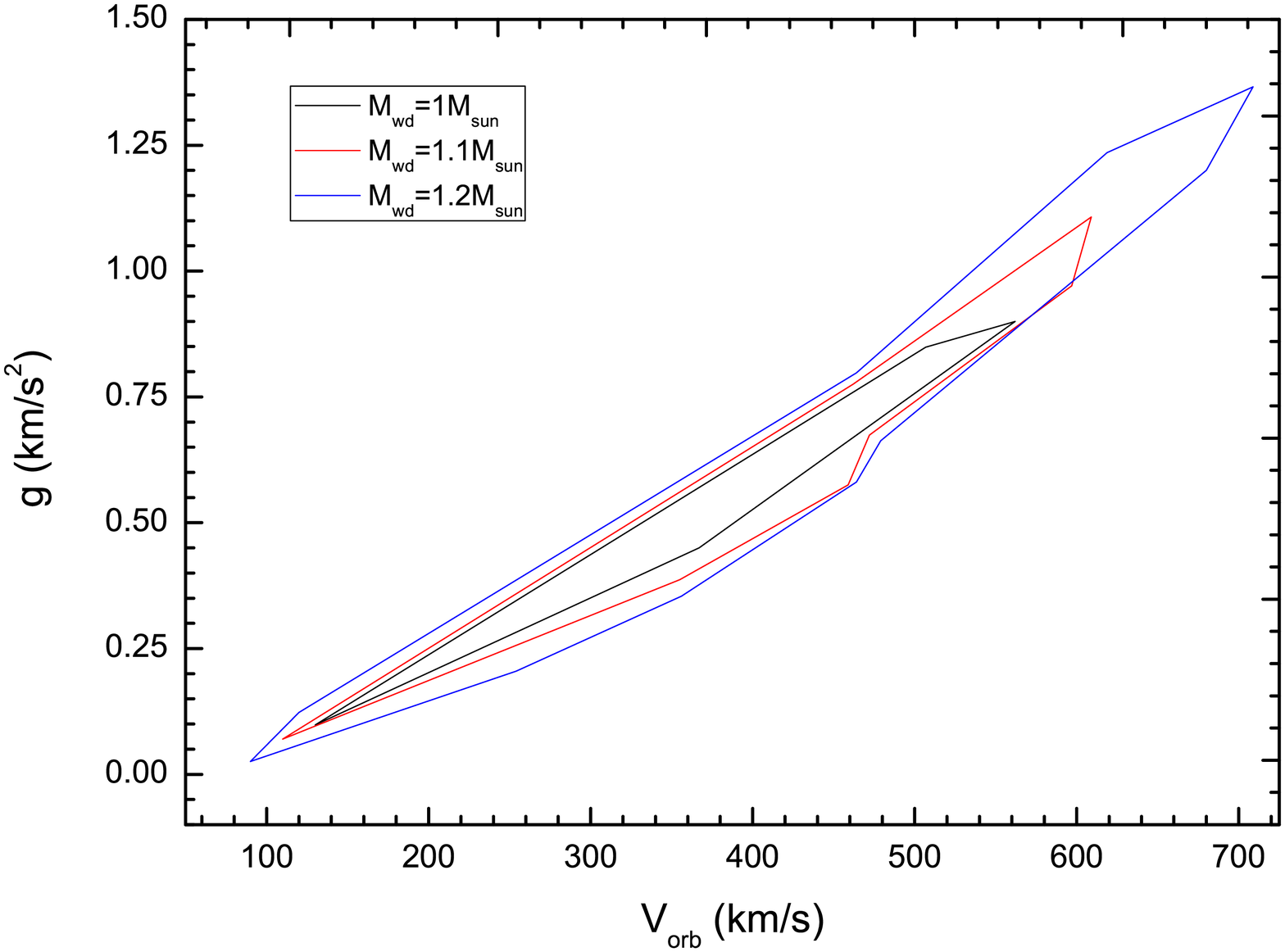}
\caption{ Distributions of the  surface gravities and orbital velocities of the companion
 when the WD grows to $1.38 M_\odot$ in Cases 1 (left) and 2 (right).}\label{fig:h5}
\end{figure}

\clearpage
\begin{figure}
\includegraphics[totalheight=3in,width=3in]{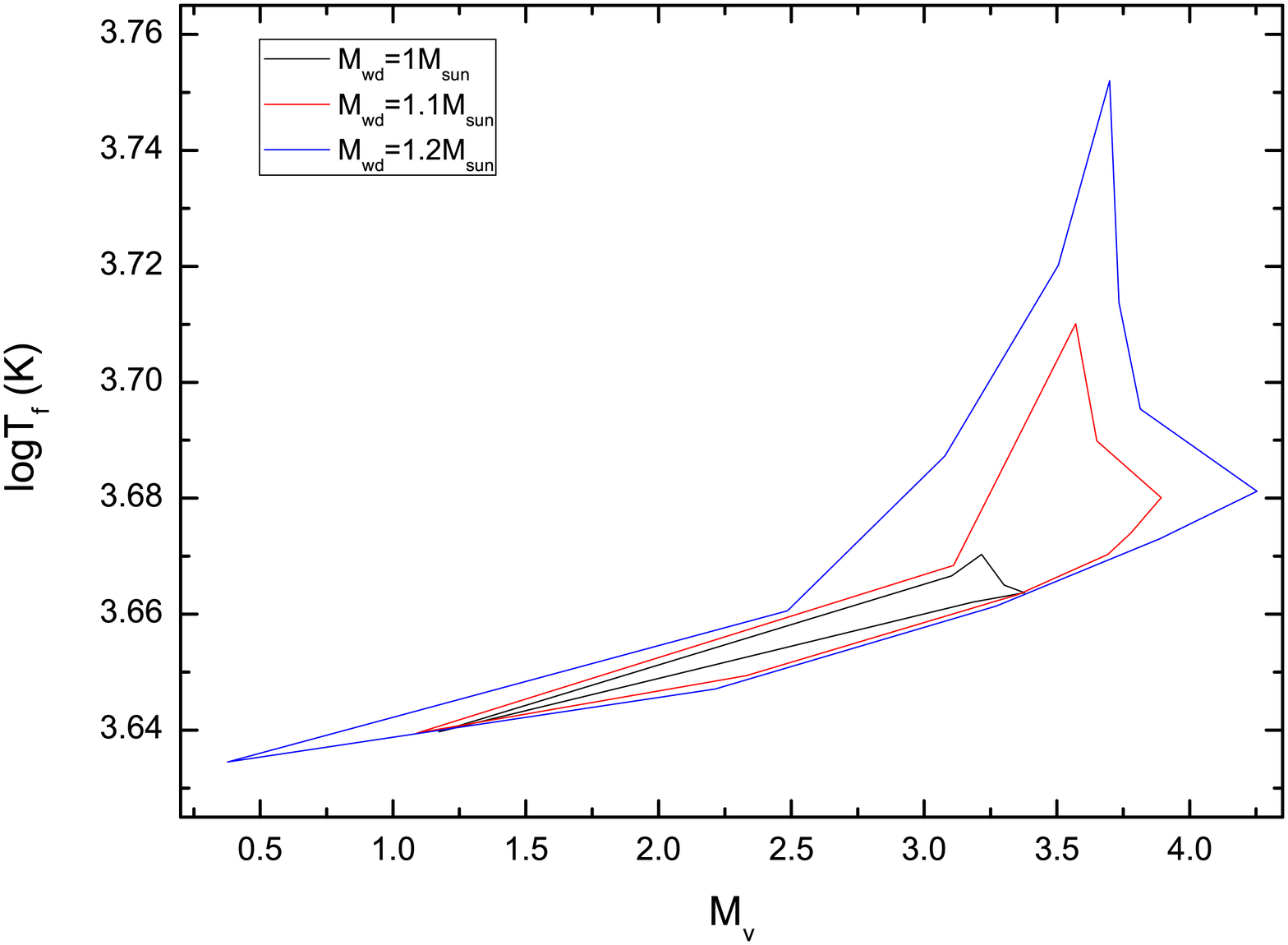}
\includegraphics[totalheight=3in,width=3in]{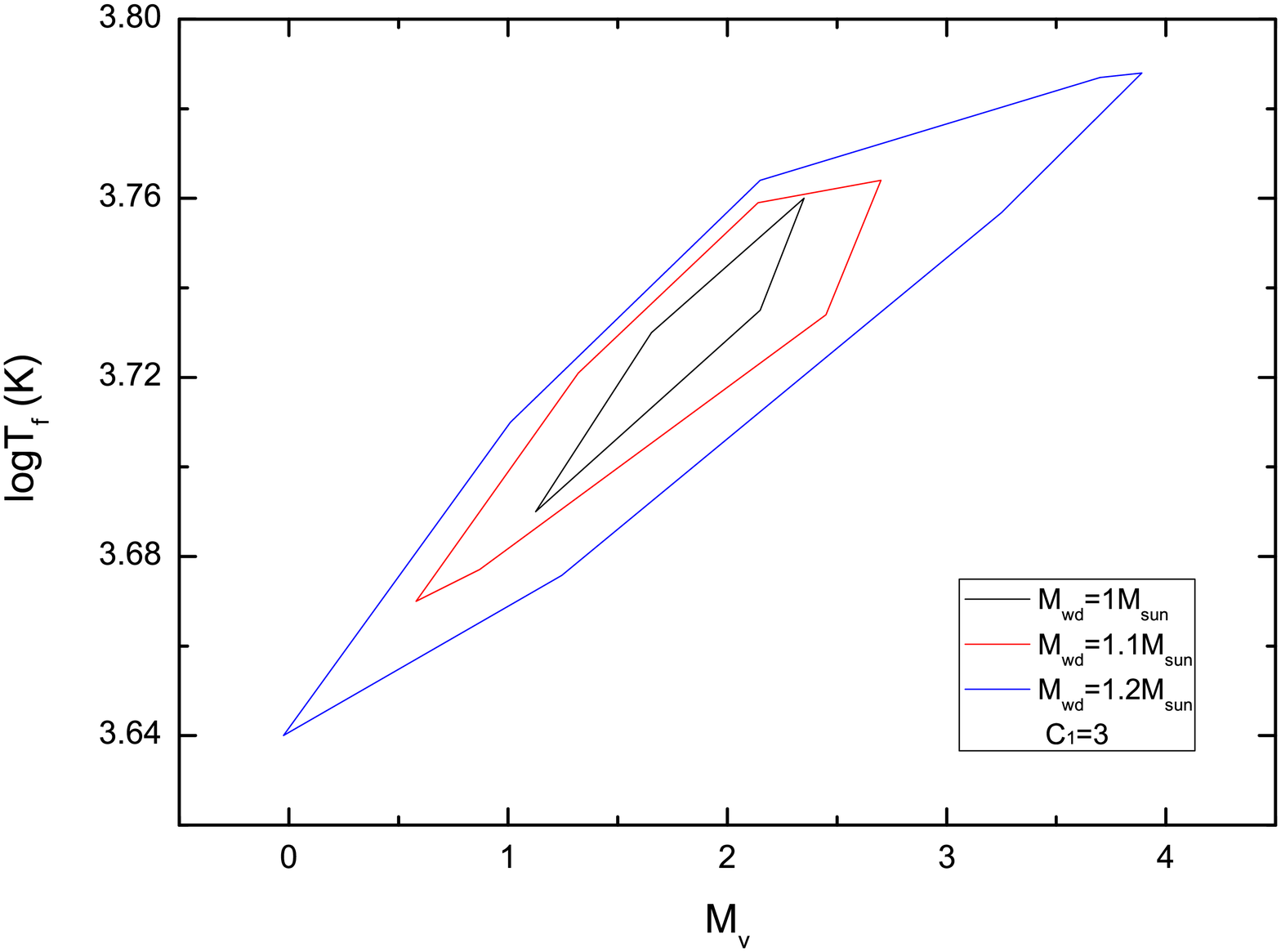}
\caption{ Distributions of the  surface effective temperatures
and absolute magnitudes of the companion when the WD grows to
$1.38 M_\odot$ in Cases 1 (left) and 2 (right).}\label{fig:h6}
\end{figure}

\clearpage

\begin{figure}
\includegraphics[totalheight=4.5in,width=5in]{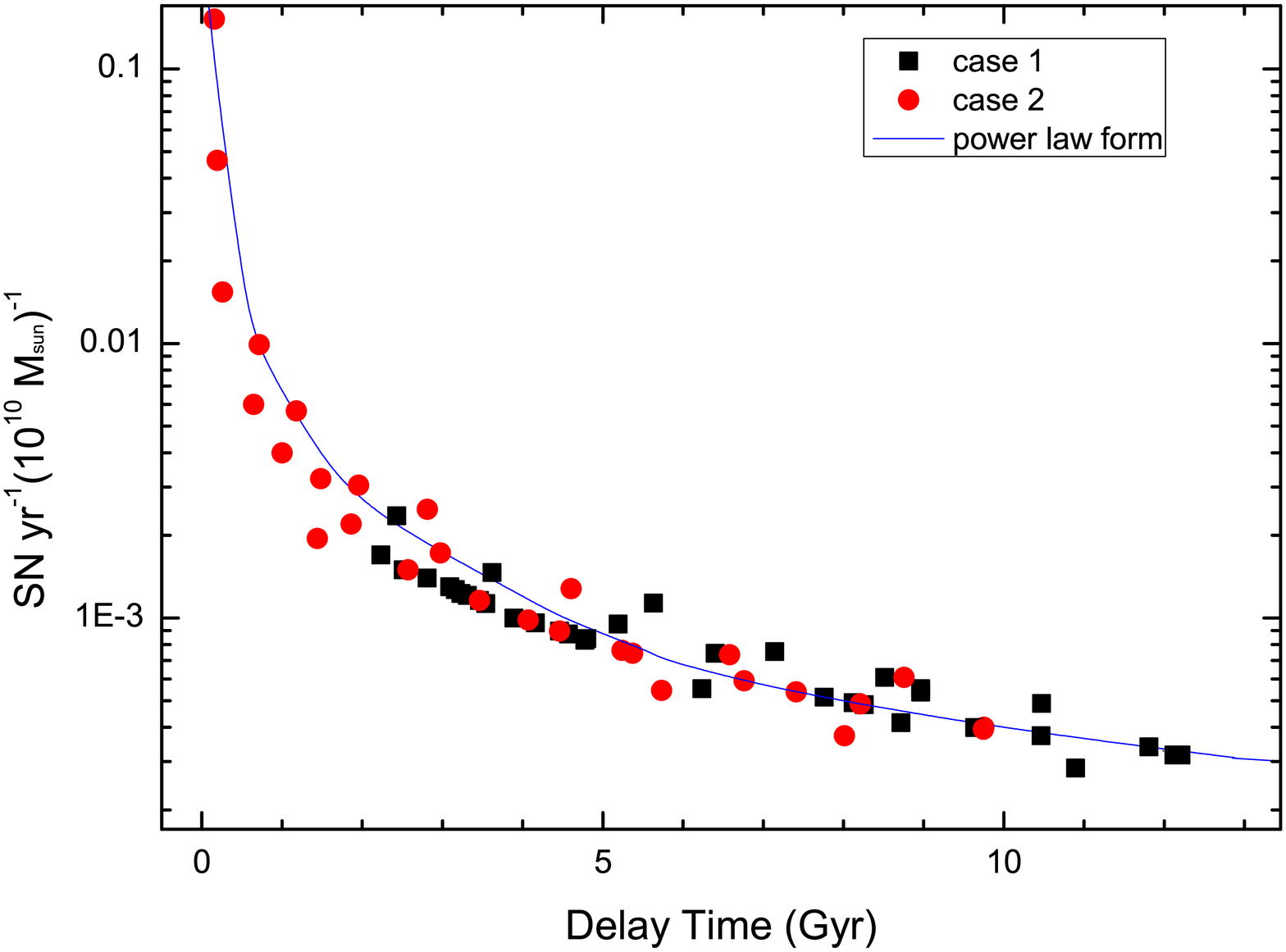}
\caption{Distributions of the delay time of SNe Ia. The blue line represents
the empirical power-law form Eq.~(7) for our Galaxy.}\label{fig:hDTD}
\end{figure}

%
%
%
%

\label{lastpage}

\end{document}